\documentstyle[prl,aps]{revtex}

\def\ll{\label}
\def\re{\ref}
\def\c{\cite}

\def\r1{(\ref{$1})}

\def\ti{\tilde}

\def\ba{\begin{array}{c}}

\def\ea{\end{array}}

\def\de{\delta}

\def\ov{\over}
\def\ha{{1\over 2}}

\def\l{\left}
\def\l({\left(}
\def\r){\right)}
\def\r{\right}

\def\be{\begin{equation}}
\def\bc{\begin{center}}
\def\ec{\end{center}}
\def\ee{\end{equation}}
\def\ed{\end{document}}
\def\bea{\begin{eqnarray}}
\def\eea{\end{eqnarray}}

\def\d{\partial}
\begin{document} 
\title{ Exact solution of Calogero model with competing  long-range
interactions}

\author{ B. Basu-Mallick$^\dagger$ and
Anjan Kundu \footnote {email: anjan@tnp.saha.ernet.in
\\ \qquad $^\dagger$biru@tnp.saha.ernet.in} \\  
  Saha Institute of Nuclear Physics,  
 Theory Group \\
 1/AF Bidhan Nagar, Calcutta 700 064, India.
 }
\maketitle
\vskip 1 cm

\begin{abstract} 

An integrable extension of the Calogero model is proposed to study the
competing effect of momentum dependent long-range interaction over the
original ${1 \ov r^2}$ interaction. The eigenvalue problem is exactly solved
and the consequences on the generalized exclusion statistics, which appears
to differ from the exchange statistics, are analyzed.  Family of `dual'
models with different coupling constants is shown to exist with same 
exclusion statistics.


\medskip
 03.65.-w,
 05.30.Pr,
 05.45.-a
\end{abstract}

\smallskip


Particles with generalized exclusion statistics (GES) introduced by Haldane
\c{haldane} is believed to play  important role in 1D non-Fermi liquids 
as well as in the edge excitations in fractional quantum Hall effect
\c{wu,wen}. Such exclusion statistics 
 can be realized microscopically in models like
 Calogero model  \c{calogero} with   ${1 \ov r^2}$  
  interaction:
\be  
H_{cs}= -  {1\over 2} \sum_{i=1}^N {\d^2 \over \d x_i^2}
+   {\omega^2\over 2}  \sum_{i=1}^N x_i^2 +
   {g \over 2} \sum_{i\neq j} {1 \over (x_i -x_j)^2}  .
\ll{cs} 
\ee
 More precisely, the GES parameter $\nu$  
  can be linked with  the coupling constant  as \c{isakov,murty,polyc} 
\be\nu = {1  \over 2} \left(1 \pm \sqrt{ 1 + 4g} \right) .
\ll{nug}\ee
On the other hand, two dimensional  anyonic model
in the lowest Landau level under strong magnetic field  can also
 be related  to  
the exclusion statistics \c{ouvry}.
However one observes  that  the GES parameter  in  
 such   anyonic model
is given by the coupling constant of a
momentum dependent long-range interaction. Similar momentum dependent
but short-range interactions   were found to appear also
 in  integrable models, like
derivative $\delta$ \c{ddelta}
 and double-$\delta$ function 
bose gases exhibiting  again  the exclusion statistics 
\c{kundu99,jakiw}.
 This motivates us to study the competing effect of a momentum dependent
Coulomb-like interaction: 
$
 H_p= \delta \sum_{i=1}^N  f_i { \d \over \d x_i }   ,
   \ \ \ \ \mbox{ where } \ \ f_i= 
    \sum_{k\neq i} {1 \over x_i - x_k}, \ 
$
over the
$1 \ov r^2$  interaction of Calogero model, 
  by adding   $H_p$ directly to Hamiltonian
  (\re{cs}). In general such additional interaction with an
independent coupling constant spoils the integrability of the original
system. However, we find that
 the proposed model
\be
 H= H_{cs} + H_p  
  ~= -  {1\over 2} \sum_{i=1}^N
 {\d^2 \over \d x_i^2} + {\omega^2\over 2}  \sum_{i=1}^N x_i^2
   +{g \over 2} \sum_{i\neq j} {1 \over (x_i -x_j)^2} + 
  \delta \sum_{i=1}^N  f_i { \d \over \d x_i }   ,
  \ll{h} 
\ee
not only retains its exact solvability 
but also allows explicit  construction of  
Lax pair and  infinite set of
conserved quantities.  Moreover, the  additional 
coupling constant 
$\de$  competes now with the coupling constant $g$ 
to modify  considerably the GES picture. In fact, as will be shown
below,
 the GES parameter becomes functionally dependent on both the coupling
constants and  remains nontrivial even in the absence of the 
 original ${1 \over r^2}$ interaction. This  would give rise to
 a family of models with   same exclusion statistics 
 but with different sets of  coupling constants lying on a parabolic curve. 
Furthermore, contrary to the Calogero model, the exchange statistics 
in the present model seems to  differ from its exclusion statistics.

To reveal the above features   
 we recall  that, Calogero like models  have
 been solved recently by mapping them to
a system of free oscillators \c{panigrahi,sogo,bf,wadati1}.
For solving  model (\re{h}) through a  similar technique, 
we conjecture  first that its ground state 
is given  by the Laughlin-type  wave function  
\be
\psi_{gr}= \prod_{i<j} (x_i-x_j)^\nu e^{
  - {\omega\over 2}  \sum_{i=1}^N x_i^2 } , 
\ll{psigr}
\ee
  $\nu$ being an  unknown parameter to be determined
later. 
Using  expression  (\re{psigr})  
 we may simplify the  Hamiltonian (\re{h}) as
\be
H_1=\psi_{gr}^{-1} \left( H-E_{gr} \right) \psi_{gr} 
  = -  {1\over 2} \sum_{i=1}^N {\d^2 \over \d x_i^2} -
  (\nu- \delta)\sum_{i=1}^N  f_i { \d \over \d x_i }  +
  \omega \sum_{i=1}^N  x_i { \d \over \d x_i } ,
\ll{H1}
\ee
where 
\be
E_{gr}= {N \omega \over 2} \left [ 1 + (N-1)(\nu -\delta ) \right ] .
\ll{Egr}\ee
For this simplification 
we have    dropped 
the $1 \ov r^2$ term by equating its coefficient to zero, which
gives $ g=\nu^2-\nu(1+2 \de) $ and
 in turn  determines the power index $\nu$ as
\be\nu = {1  \over 2} \left((1+2\delta ) \pm \sqrt{ (1+ 2\delta )^2 + 4g}
\right) .
\ll{nu}\ee
We notice that the parameter
$\nu,$ which is linked to the symmetry of the wave function
 (\re{psigr})  and  related to the exchange statistics,   
  depends  on both the coupling
constants of  model (\re{h}) in an involved way.

We observe further that  the transformed 
 Hamiltonian (\re{H1}) may be  expressed  as 
$
H_1= S^- + \omega S^3,
$
 with the Lassalle operator $S^-=
   -  {1\over 2} \sum_{i=1}^N {\d^2 \over \d x_i^2} -
(\nu - \delta )\sum_{i=1}^N  f_i { \d \over \d x_i } \ \ $ and  the Euler
operator $S^3= \sum_{i=1}^N  x_i { \d \over \d x_i }, $
 satisfying  the commutation relation 
$\ \ [S^-,S^3]=2 S^-$ \c{wadati1}.
  Using therefore the well known Baker-Hausdorff transformation 
we can remove the $S^-$ part from  $H_1$
and through  some additional similarity 
 transformations  reduce it finally to a free oscillator model
\be
H_2 = {\cal S}^{-1} H_1 {\cal S}
  = -  {1\over 2} \sum_{i=1}^N {\d^2 \over \d x_i^2} +
 {\omega^2 \over 2}  \sum_{i=1}^N x_i^2 - {\omega N \over 2},
\ll{H2}
\ee
where ${\cal S} =e^{ {1\over 2 \omega } S^- } 
e^{ {1\over 4 \omega } \nabla^2 } 
e^{ {\omega \over 2} \sum_{i=1}^N x_i^2 }  $ and 
$ \nabla^2 = \sum_{i=1}^N {\d^2 \over \d x_i^2} $.
This mapping  
 allows us to write the exact 
 eigenfunctions for  the extended Calogero model (\re{h})
as
\be \psi_{n_1n_2\cdots n_N}=\psi_{gr} \ {\cal S}
\psi^0_{n_1n_2\cdots n_N}.\ll{psi} \ee
Here  the excitation numbers $n_i$'s  are positive integers
 obeying bosonic selection rule
 $n_i \leq n_{i+1}$ and
 $\psi^0_{n_1n_2\cdots n_N}$ corresponds to the symmetrized form 
of the  free oscillator eigenfunction, which in turn is given
   through the product of  Hermite polynomials.
The symmetrization of the wave function
with respect to  particle coordinates, as was implemented in Calogero model
\c{panigrahi}, is also needed here  to avoid singularity
 in the normalization of the wave function
(\re{psi}). Therefore, in spite of the fact that the interacting Hamiltonian
 (\re{h})
is convertible to the free oscillator model, the need for symmetrization 
shows that the many-particle correlation is in fact inherent in this model.  

The eigenvalues of the Hamiltonian (\re{h})
 corresponding to the states  (\re{psi}) will  
  naturally be given by 
\be 
E_{n_1,n_2, \cdots , n_N} = E_{gr}+\omega \sum_{i=1}^N n_i= 
 {N \omega \over 2} \left [ 1 + (N-1)(\nu -\delta ) \right ] +
 \omega \sum_{i=1}^N n_i .
\ll{e}
\ee
It is evident that for all $n_i=0,$  
the energy $E_{n_1,n_2, \cdots , n_N}$
  attains its minimum value   
$E_{gr}$. At the same time, as can be easily worked out
 from Eqn. (\re{psi}),
 the corresponding  eigenfunction  reduces to 
$\psi_{gr}$ (\re{psigr}). 
This proves that $\psi_{gr}$ is indeed the ground state wave function 
for Hamiltonian (\re{h}) with eigenvalue $E_{gr}$. 
For exploring the exclusion statistics  
 of this model, 
we observe  that 
its energy spectrum  (\re{e}) can be expressed  
 exactly in the form of free oscillator:  
$E_{n_1,n_2, \cdots , n_N} = 
 {N \omega \over 2}  + \omega 
\sum_{i=1}^N \bar n_i, 
$
 where $
 \bar n_i= n_i+ (\nu -\de)(i-1) 
$ are quasi-excitation numbers.  
  However such numbers are no longer integers and satisfy a  modified 
 selection rule
\be  
\bar n_{i+1} -\bar n_i \geq (\nu-\de),
\ll{excl} \ee
 which clearly restricts the difference
between the quasi-excitation numbers to be at least  
$\tilde \nu=\nu-\de$ apart.
As a consequence the extended
 Calogero model (\re{h}) exhibits  GES property with  parameter
\be \tilde \nu \equiv
\tilde \nu (\de,g)= \ha\left( 1 \pm \sqrt{ (1+ 2\delta )^2 + 4g} \right),
 \ll{tnu} \ee
 which is clearly a function of  both the  
  coupling constants of the model.
It is obvious    that for $\de \neq 0$, the GES parameter $\tilde \nu $
  is different from the power 
index $ \nu $ (\re {nu}), which is responsible for the 
symmetry of the wave function.
Therefore one may conclude that unlike  Calogero model,
 the exclusion statistics for  model
(\re{h}) differs from its exchange statistics.
This might lead to unusual  situations, where the particles with bosonic 
exclusion 
statistics (with $\ti \nu=0$) might have nonsymmetric wave function 
(with $ \de=\nu \not=0$) and reversely, symmetric wave function 
(with $ \nu =0$) might correspond to   fermionic particles 
(with $ \ti \nu =1=-\de$) or those with fractional exclusion statistics
(with $ -\de=\ti \nu \not=0$).

It is evident that equation  (\re{tnu})
for fixed values of $\ti \nu$ describes  a family of parabolic
curves  in the coupling constant plane $(\de, g).$ 
As a consequence of this, the competing effect of the independent
coupling constants $g$ and $\de$  can  make
   the GES feature of  (\re{h}) much
  richer in comparison with the  Calogero model.
For example, while
bosonic (fermionic) excitations in the Calogero model
 occur only in the absence of long-range interaction,
 the quasi-excitations in  (\re{h}) can behave as 
pure bosons (fermions)  
 even in the presence of both the long-range
  interactions satisfying the constraint
$\ti \nu(\de,g)=0$ \ ($\ti \nu(\de,g)=1$). Both of these constraints
 lead to 
the same  parabolic curve $g=-\de(1+\de)$. 
A family  of such parabolas with shifted  apex points are generated 
 for other values of 
$\ti \nu$ and the lowest apex point is attained at $\ti \nu=\ha$,
where
the quasi-excitations would behave as semions.
Note that all points
  along any of these  parabolic  curves
  would correspond to 
 different Hamiltonians having 
 different sets of coupling constants. However all such models 
 will have the same GES parameter $\tilde \nu$  
 and  hence generate  the same energy spectrum. Therefore such
 models   may 
 be considered as `dual' to each other.
For example,   the model
\be
 H_{\de}  
  ~= -  {1\over 2} \sum_{i=1}^N {\d^2 \over \d x_i^2} 
+ {\omega^2\over 2}  \sum_{i=1}^N x_i^2  +
  \delta \sum_{i=1}^N  f_i { \d \over \d x_i }   , \ll{hde} 
\ee
 will be dual to the  Calogero model (\re{cs})
 with coupling constant $g=\de(\de+1).$

As is well known, in  Calogero model
the same  coupling constant $g$
may correspond to  two distinct GES parameters
 related by a   symmetry:  
 $ \nu \to 1- \nu$.
To reveal such a reflection
symmetry   in the present case together  with some other details,
we may look into the projection of 
 the surface described by  equation (\re{tnu}) on the 
$(\de, \ti \nu)$ plane for fixed values of $g$. This   
  gives a  family of 
 two-branched hyperbolas, which
reduce to their asymptotes at $g=0$.
When  $g$ changes from positive to negative values, the orientation of the
hyperbolas  flips by an angle $ {\pi
\ov 2}$.
A remarkable point is that, while in the Calogero model the minimum
allowed value of the coupling constant $g$ is $-{1\ov 4}$,
 the corresponding value of $g$ 
 in the present model  is $-{1\ov 4}(1+2 \de)^2$,
which has no lower bound.
Note further that to avoid singularity in the wave function (\re{psi})
 the power index $\nu $ should be positive and at the same time 
 for physical reasons the GES
parameter $\ti \nu$ should not be negative.
Consequently, for $g > 0$  only  the upper-branch  hyperbolas, while
 for $g<0$ a segment of the right-branch  hyperbolas  survive
(as shown in Fig. 1).
As is evident from the figure, 
  along any allowed  segment  with fixed $g<0, $ 
 there exists certain domain where
$\ti \nu$  becomes a
 double-valued  function of $\de.$
From Eqn. (\re{tnu}) we can explicitly
 calculate this domain as $\ \sqrt{-g} -0.5 \leq \de \leq \sqrt{0.25-g} -0.5
$.
Therefore the same set of coupling constants $g$
and $
\de$ may lead to two different values of the GES parameters
 revealing the  reflection symmetry:  
 $\ti \nu \to 1-\ti \nu$ in model (\re{h}) in a more involved way.
It may also be seen from the figure that
 on any hyperbolic curve with fixed $g>0$,  
 two different values of  $\de: \de_\pm=-\ha\pm x$,
  give the same $\ti \nu$ for any $x$. 
As a consequence
 the   models with the coupling constants $(g,\de_-)$
and $(g,\de_+)$ become  dual to each other for any   positive  $g$.
In particular, the  Calogero model (\re{cs})
would be dual to 
\be
H_{cs}^{(dual)}  
  ~= -  {1\over 2} \sum_{i=1}^N {\d^2 \over \d x_i^2} 
+ {\omega^2\over 2}  \sum_{i=1}^N x_i^2   +
   {g \over 2} \sum_{i\neq j} {1 \over (x_i -x_j)^2} - 
   \sum_{i=1}^N  f_i { \d \over \d x_i } . 
\ll{dual}\ee

We  show now that similar to  Calogero model, its extension
(\re{h}) also represents a quantum integrable system with
 required number of conserved quantities.
The   Lax operator of this system
 is a $N \times N$ matrix with  elements 
\be
L_{jk}= i \ ( - {\d \over \d x_j} + \de f_j ) \  \de_{jk} + i \ {\ti \nu}
{ (1- \de_{jk}) \over x_j - x_k},
\ll{L}\ee
which reproduces the $L$ operator of Calogero model at $\de \rightarrow 0$
limit. 
The complementary Lax operator $M$ on the other hand 
 has the same form  as in  
Calogero model  \c{wadati2}, but   with its  parameter $\nu$ replaced by $ 
\ti \nu$. This pair of operators 
satisfy the Lax equation $[H,L^\pm]=[L^\pm ,M]\pm \omega L^\pm$,
 where $L^\pm=L\pm \omega Q$ with
$Q_{jk}=i \ x_j \de_{jk}$.
    This Lax  equation  leads to 
the set of conserved quantities 
$I_m= \sum_{jk}\left( (L^+ L^-)^m \right)_{jk}, \ m=1,2,3, \ldots ,$ 
ensuring the integrability of the system.
Note that the  Hamiltonian (\re{h}) of the extended  model  
  coincides with the  conserved quantity ${I_1 \over 2}+ E_{gr}$. 

We would like to point out here that the proposed Hamiltonian 
(\re{h}) lacks hermiticity property when $\delta $ and $g$ are real 
parameters. However, we can easily construct another variant 
of this model by adding a momentum dependent interaction like 
$H_p^{'} = {{\tilde \delta} \over 2} 
\sum_{j=1}^N (f_j p_j + p_j f_j) ,$ with $p_j = -i {\d \over \d x_j},$
to the original Calogero model. This clearly leads to a hermitian 
 Hamiltonian of the form 
\be
 H ~= -  {1\over 2} \sum_{i=1}^N
 {\d^2 \over \d x_i^2} + {\omega^2\over 2}  \sum_{i=1}^N x_i^2
   +{{\tilde g} \over 2} \sum_{i\neq j} {1 \over (x_i -x_j)^2} -i 
  {{\tilde \delta } \over 2} \sum_{j=1}^N \left (
  f_j { \d \over \d x_j }  +   { \d \over \d x_j } f_j \right ) ,
  \ll{hh} 
\ee
where $ {\tilde g} $ and $ {\tilde \delta } $ are real parameters.
It is remarkable that (\re{hh}) can  be expressed again  in
 the form (\re{h}),
where the coupling constants of the model are given by 
\be 
g = {\tilde g} + i {\tilde \delta } , ~~
 \delta =  - i {\tilde \delta } .
  \ll{t} 
\ee
Therefore all the procedures followed above for solving the eigenvalue 
problem of (\re{h}) go through exactly in the same  way 
for the hermitian variant (\re{hh}). This however leads to some
 subtleties.  In particular, from (\re{nu}) we see that 
the `exchange statistics' parameter 
  $\nu$ becomes complex  valued in this case and its 
real and imaginary parts can be
 expressed through the coupling constants  as
\bea 
\nu &=&\nu_R + i \nu_I,  \nonumber \\
\nu_R& =& { 1 \over 2}\left(1\pm \sqrt{ 1 + 4 ({\tilde g} - {\tilde \delta }^2 )
}\right)  ,
~~   \nu_I = - {\tilde \delta } ,
\ll{ri}
\eea
for the  choice 
 $~~ {\tilde g} \geq   {\tilde \delta }^2 - {1\over 4} $.
Combining   relations (\ref{t}) and (\ref{ri})
we get $\tilde \nu =\nu-\delta=\nu_R, $ i.e.  the exclusion
statistics parameter $\tilde \nu$ for the hermitian model (\re{hh}) is
given by the real part of the exchange statistics parameter.
As a consequence, the energy spectrum in this case will be given by
$~~ 
E_{n_1,n_2, \cdots , n_N} = E_{gr}+\omega \sum_{i=1}^N n_i= 
 {N \omega \over 2} \left [ 1 + (N-1)\nu_R  \right ] +
 \omega \sum_{i=1}^N n_i .
~~$
On the other hand both the real and imaginary parts of $\nu$ appear in the 
ground state wave function 
\be
\tilde \psi_{gr}= \prod_{i<j} (x_i-x_j)^{-i \tilde \delta}\psi^0_{gr} ,
\ll{psigr0}
\ee
where 
$
\psi^0_{gr}= \prod_{i<j} (x_i-x_j)^\nu_R  e^{
  - {\omega\over 2}  \sum_{i=1}^N x_i^2 } . $ Note that   
 $\nu_R$ depends  on the redefined coupling constant $g^{'}=
{\tilde g} - {\tilde \delta }^2$
  exactly in the same way as   $\nu$  
depends on its coupling constant  $g$  given by
(\ref{nug}) in case of Calogero model.  
 And since    $\nu-\delta$ is  replaced  here by $\nu_R$,  
 the operator $S^-$ constructed above is also  reduced to the corresponding
operator for the Calogero model with redefined coupling constant.
Consequently the wavefunctions for the hermitian model (\ref {hh})
and the Calogero model will have similar structures.
Moreover, since the complex part of the wavefunctions appearing due to
 (\re{psigr0})
 does not affect the normalization, the normalizability
of the eigenfunctions associated with the Hamiltonian (\ref{hh})
 can be easily established  by  following the same
argument as in the case of  Calogero model.

Thus we have established that,  additional long-range interactions in 
 Calogero model not only  preserves its integrable structure leading
to exact eigenvalue solutions, but also shows significantly richer
exclusion statistics properties. Due to 
the competing effect of two different long-range interactions, one can
 construct  a family of dual models with different sets of
coupling constants, but having same
exclusion statistics and  same energy eigenvalues.
 More importantly, the non-coincidence of the
exchange statistics with the exclusion statistics, a possibility in
one-dimension, is explicitly realized in such models.
As a consequence  the dual models themselves differ in their exchange
statistics.    
The inclusion of spin degrees of freedom and periodic boundary
condition in the proposed model could be an important problem
for future investigation \c{bk}.

\vskip 2cm

\input {figure1.tex}
\vskip 1cm
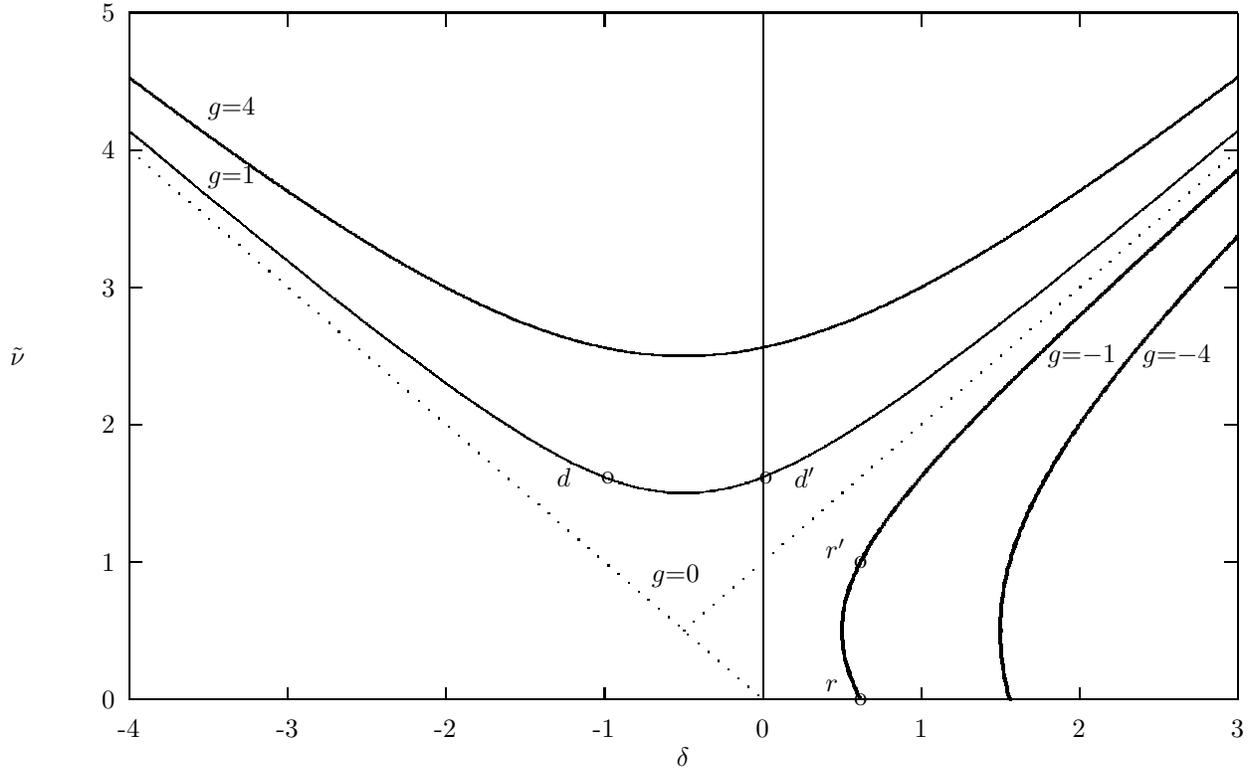
{\bf Figure 1}: Plot of the exclusion statistics parameter
${\tilde \nu}$ as a function of $\de $ for fixed values of
$g=\pm1,\pm4$. The points $d,d'$  correspond to dual models (\re{dual}) and
(\re{cs}) with same
$\ti \nu = {\sqrt 5 +1\over 2}.$  The points $r,r'$ 
 with $\de={\sqrt 5 -1\over 2}$ are related by
reflection symmetry. 

\end{document}